\documentstyle[twoside,fleqn,espcrc2,rotate,epsf]{article}
\newcommand{\IM}{\mbox{\rm Im}}
\newcommand{\nn}{\nonumber}

\newcommand{\as}{\alpha_{s}}
\newcommand{\Mn}{{\cal M}_n}
\newcommand{\ve}{\varepsilon}
\newcommand{\eqn}[1]{(\ref{#1})}

\newcommand{\gev}{\mbox{\rm GeV}}
\newcommand{\MSb}{{\overline{MS}}}

\newcommand{\gsim}{~{}_{\textstyle\sim}^{\textstyle >}~}
\newcommand{\lsim}{~{}_{\textstyle\sim}^{\textstyle <}~}

\title{
\vspace{-2.6cm}
{\small\sf
\rightline{HD-THEP-98-50}
\rightline{FTUV/98-75}
\rightline{IFIC/98-76}}
\bigskip
{\bf Bottom quark mass from QCD sum rules for the $\Upsilon$ system}}

\author{M. Jamin\address{
       {\em Institut f\"ur Theoretische Physik, Universit\"at Heidelberg,} \\
       {\em Philosophenweg 16, D-69120 Heidelberg, Germany}}
        and
        A. Pich\address{
       {\em Departament de F\'{\i}sica Te\`orica, IFIC,
            CSIC --- Universitat de Val\`encia,} \\
       {\em Dr. Moliner 50, E--46100 Burjassot, Val\`encia, Spain}
\protect\thanks{Invited talk at the Euroconference on Quantum Chromodynamics
        (QCD'98), Montpellier, July 1998}}}

\begin{document}

\begin{abstract}
The talk presents an update of the bottom quark mass determination from
QCD moment sum rules for the $\Upsilon$ system by the authors \cite{jp:97}.
Employing the $\MSb$ scheme, we find $m_b(m_b) = 4.19 \pm 0.06\,\gev$.
The differences to our previous analysis will be discussed and we
comment on the determination of the pole mass for the bottom quark.
\end{abstract}

\maketitle


\section{Introduction}

A precise determination of the bottom quark mass, being one of the
fundamental parameters in the Standard Model (SM), is of paramount
interest in several areas of present day particle phenomenology.
In the past, QCD moment sum rule analyses have been successfully applied
for extracting charm and bottom quark masses from experimental data on the
charmonium and bottomium systems respectively \cite{svz:79,rry:85,nar:89}.

The basic ingredient in these investigations is the vacuum polarisation
$\Pi(q^2)$,
\begin{displaymath}
\left(q_\mu q_\nu-g_{\mu\nu}q^2\right)\Pi(q^2) = i \!\!\int \!\! dx \,
e^{iqx} \langle  T\{j_\mu(x) j_\nu(0)\} \rangle
\end{displaymath}
induced by the vector current $j_\mu\equiv\bar Q\gamma_\mu Q$ for a heavy
quark $Q$. Inclusive quantities, like $\Pi(q^2)$, where no reference to
a particular hadronic state is needed, are especially well suited for a
description in terms of quarks and gluons. Theoretically, $\Pi(q^2)$ can
be calculated in the framework of the operator product expansion (OPE).

On the phenomenological side, the imaginary part of $\Pi(q^2)$ is a physical
quantity, related to the experimentally measured cross section for
$e^+e^-\!\to Q \bar Q$:
\begin{displaymath}
R(s) \equiv \frac{1}{Q_Q^2}
\frac{\sigma(e^+e^-\!\to Q \bar Q)}{\sigma(e^+e^-\!\to\mu^+\mu^-)} =
12\pi \,\IM\,\Pi(s\!+\!i\ve) .
\end{displaymath}
Because of its analytic properties, $\Pi(s)$ satisfies a dispersion relation.
Taking $n$ derivatives of $\Pi(s)$ at $s=0$, one arrives at the following
expression for the $n$th moment $\Mn$,
\begin{eqnarray}
\label{eq:Mn}
\Mn &\!\!\! \equiv &\!\!\! \frac{12\pi^2}{n!}\left.\left(4M^2\,\frac{d}{ds}
\right)^n \Pi(s)\right\vert_{s=0} \\ & &
\hspace{-12mm} = \, (4M^2)^n\!\int\limits_0^\infty\!\!ds\,\frac{R(s)}{s^{n+1}}
\, = \,  2 \!\int\limits_0^1 \!\! dv \, v(1-v^2)^{n-1} R(v) \nn
\end{eqnarray}
where $v=\sqrt{1-4M^2/s}$. $M$ is the heavy quark pole mass, defined as
the pole of the perturbatively renormalised propagator.

Under the assumption of quark-hadron duality, the moments $\Mn$ can either
be calculated theoretically in renormalisation group improved perturbation
theory, including non-perturbative condensate contributions, or can be
obtained from experiment. In this way, hadronic quantities like masses
and decay widths get related to the QCD parameters $\alpha_s$, quark masses
and condensates.

As can be seen from eq.~\eqn{eq:Mn}, for large values of $n$ the moments
become more and more sensitive to the quark mass, but also to the threshold
region which is dominated by large Coulombic corrections growing like
$(\as\sqrt{n})^k$. The situation is greatly improved by utilising the quark
mass in the $\MSb$ scheme where perturbative QCD corrections turn out to be
much smaller than in the pole mass scheme. In addition for large $n$ the
non-perturbative terms, the first being the gluon condensate, start to
dominate over the perturbative contribution. It will prove possible to find
an intermediate range of $n$ where the sum rules are already very sensitive
to the bottom quark mass and contributions from higher states are sufficiently
suppressed, whereas the corrections from higher order as well as
non-perturbative corrections are still under control.

\section{Perturbative spectral function}

The perturbative spectral function $R(v)$ can be expanded in powers of the
strong coupling constant $\as$,
\begin{displaymath}
R(v) = R^{(0)}(v) + a\,R^{(1)}(v) + a^2 R^{(2)}(v) \,,
\end{displaymath}
with $a\equiv \as/\pi$. From this expression the corresponding moments
$\Mn^{(k)}$ can be calculated via the integral of eq.~\eqn{eq:Mn}. The
first two terms are known analytically and can for example be found in
ref.~\cite{jp:97}.

The second order contribution is as yet not fully known analytically.
However, exploiting available results at high energies, analytical
expressions for the first eight moments $\Mn^{(2)}$ ($n=1,\ldots,8$)
and the known threshold behaviour of $R^{(2)}(v)$, Pad\'e approximants
for the function $R^{(2)}(v)$ could be extracted \cite{cks:96}.
This information is sufficient for a reliable determination of $\Mn^{(2)}$
for $n$ in the region of interest.

Until now, our expressions for the moments have been written in terms of
the pole quark mass $M$. The relation between the pole and $\MSb$ mass
is known up to second order in $\as$ \cite{gbgs:90}:
\begin{displaymath}
m(\mu_m) = M[ 1 + a(\mu_a)\,r_m^{(1)}(\mu_m) + a^2 r_m^{(2)}(\mu_a,\mu_m) ]
\end{displaymath}
where $\mu_a$ and $\mu_m$ are renormalisation scales for coupling and mass
respectively. Explicit expressions for $r_m^{(1)}$ and $r_m^{(2)}$ can be
found in appendix~A of ref.~\cite{jp:97}. Using this relation one finds
the following expression for the moments $\Mn^{(1)}$ and $\Mn^{(2)}$ in
the $\MSb$ scheme:
\begin{eqnarray}
\label{eq:3.9}
\overline\Mn^{(1)} &\!\! = &\!\! \Mn^{(1)} + 2n\,r_m^{(1)}\Mn^{(0)}
\,, \nn \\
\overline\Mn^{(2)} &\!\! = &\!\! \Mn^{(2)} + 2n\,r_m^{(1)}\Mn^{(1)}
\nn \\ & & + \, n\Big( 2r_m^{(2)}+(2n-1)r_m^{(1)^2} \Big)\Mn^{(0)} \nn \,.
\end{eqnarray}

These expressions for the moments will be used in our numerical analysis
presented in section~5.

\section{Power Corrections}

The leading non-perturbative correction is the gluon-condensate contribution
to the massive vector correlator. It is known at the next-to-leading order
\cite{bro:94}. The relative growth of ${\cal M}_{n,G^2}^{(0)}/\Mn^{(0)}$ is
proportional to $n^3$. Therefore, the non-perturbative contribution grows
much faster than the perturbative moments. In addition, in the pole mass
scheme, the next-to-leading order correction to ${\cal M}_{n,G^2}$ is of the
same size or larger as the leading term. Because the perturbative expansion
for the gluon condensate cannot be trusted, we shall restrict our analysis to
the range $n\leq20$ where its contribution to the $b\bar b$ moments is
below 3\%. For different reasons to be discussed below, we shall anyhow
restrict the analysis to moments $n\lsim 15$. Thus the gluon condensate
can be safely neglected.

\section{Phenomenological spectral function}

The first six $\Upsilon$ resonances have been observed. Their masses are
known rather accurately, and their electronic widths have been measured with
an accuracy which ranges from 3\% for the $\Upsilon(1S)$ to 23\% for the
$\Upsilon(6S)$. For our purposes, the narrow--width approximation provides
a very good description of these states, because the full widths of the first
three $\Upsilon$ resonances are roughly a factor $10^{-5}$ smaller than the
corresponding masses, and the higher-resonance contributions to the moments
are suppressed:
\begin{equation}
\label{eq:phen}
{\overline\Mn\over (4m_b^2)^n} = \frac{9\pi}{\bar\alpha^2 Q_b^2}\,
\sum\limits_{k=1}^6 \frac{\Gamma_{kS}}{M_{kS}^{2n+1}} +
\int\limits_{s_0}^\infty \!ds \,\frac{R(s)}{s^{n+1}} \,,
\end{equation} 
where $\Gamma_{kS}\equiv\Gamma[\Upsilon(kS)\to e^+e^-]$ and the running
fine structure constant $\bar\alpha^2=1.07\,\alpha^2$ \cite{pdg:98}.

The $e^+e^-\!\to b\bar b$ cross-section above threshold is unfortunately very
badly measured \cite{cleo:91}. The second term in Eq.~\eqn{eq:phen} accounts
for the contributions to $R_b$ above the sixth resonance as well as additional
continuum contributions between $\Upsilon(4s)$ and $\Upsilon(6s)$ and is
approximated by the perturbative QCD continuum. Hence the threshold
$M_0\equiv\sqrt{s_0}$ should lie somewhat lower than the mass of the next
resonance which has been estimated in potential models. It is found to be 
$M_{\Upsilon(7s)}\approx 11.2\,\gev$. A more precise estimate of $M_0$ can
be obtained by requiring stability of the sum rule for very low moments.
From the numerical analysis to be discussed below, we find
$M_0\approx 11.1\,\gev$. On the other hand only taking into account the
first three $\Upsilon$ resonances and requiring stability of the sum rule
we obtain $M_0\approx 10.6\,\gev$ as expected.

\section{Numerical analysis}

Using the theoretical expression for the moments of eq.~\eqn{eq:Mn} and
the phenomenological Ansatz \eqn{eq:phen}, the bottom quark mass
can be calculated as a function of $n$. This is shown in figure~1 for
various values of the input parameters.
\begin{figure}[thb]
\vspace{-6mm}
\centerline{
\rotate[r]{
\epsfxsize=2in
\epsffile{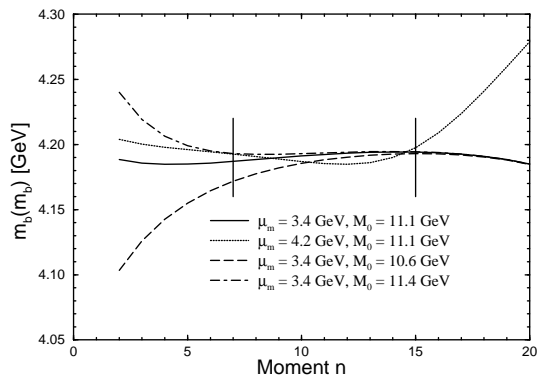} }}
\vspace{-10mm}
\caption[]{The bottom quark mass $m_b(m_b)$ as a function of $n$.
(For detailed description see text.) \label{fig:1}}
\vspace{-10mm}
\end{figure}

In our previous work \cite{jp:97}, besides $m_b$, we have also performed
a determination of $\as(m_b)$ from the $\Upsilon$ sum rule. It was
demonstrated that upon evolution to the scale $M_Z$ the value of $\as(M_Z)$
was completely compatible with current world averages though the error
was larger. Thus, in this work we now decided to use $\as$ as an input
parameter within the current average $\as(M_Z)=0.119\pm0.002$ \cite{pdg:98}.
The uncertainty in $\as$ is included in the final error on $m_b$.

Although in the $\MSb$ scheme the higher order perturbative corrections at a
scale $\mu_m=m_b$ are much smaller than in the pole mass scheme, nevertheless
for large $n$ they become dominant. For our central estimate of $m_b$ we
thus followed the same strategy as in ref.~\cite{jp:97}. We choose a mass
renormalisation scale $\mu_m$ where the perturbative corrections are
minimal in the range of interest for $n$. This is realised at
$\mu_m\approx 3.4\,\gev$. For this choice from the solid curve in fig.~1
it is clear that the prediction for $m_b$ is stable even for $n\gsim 20$
where the non-perturbative contributions start to become important.

The more `natural' choice $\mu_m=m_b$ is displayed in the dotted curve.
In this case the result for $m_b$ is stable up to $n\approx 15$ and then
becomes unreliable due to $\as$ corrections growing large. For this
reason we have restricted the range of $n$ from which we determine the
bottom quark mass to $7\leq n\leq 15$. Including all variations of the
input parameters as well as the renormalisation scales in our error
estimate~\cite{jp:97}, as our final result for the running $\MSb$ mass
$m_b(m_b)$ we obtain:
\begin{displaymath}
\label{eq:mbmb}
m_b(m_b) \; = \; 4.19 \pm 0.06\,\gev \,.
\end{displaymath}

For very low moments the continuum contribution in eq.~\eqn{eq:phen} is
not very well suppressed and thus $m_b$ displays a strong dependence on
the threshold $M_0$. As has been already discussed in the last section,
by requiring stability of the sum rule we obtain $M_0\approx 11.1\,\gev$
(solid line). For comparison, as extreme cases, in fig.~1 we have also plotted
the sum rule for the choices $M_0\approx 10.6\,\gev$ (dashed line) and
$M_0\approx 11.4\,\gev$ (dot-dashed line). 

The first case corresponds to using the perturbative QCD continuum above
the $\Upsilon(4S)$ but nevertheless including the higher resonance
contribution. This is certainly overcounting the continuum contribution.
The second choice would assume a very high mass for the $\Upsilon(7S)$
and no additional continuum contribution above the $\Upsilon(4S)$. This
is definitely on the low side of the estimates. As can be seen in fig.~1,
even for these extreme cases using moments with $n \gsim 7$ sufficiently
suppresses the dependence on the continuum contribution and gives a good
determination of $m_b$.

\section{Discussion}

Our updated result for the bottom quark $\MSb$ mass $m_b(m_b)$ as obtained
from the sum rules for the $\Upsilon$ system is in very good agreement to
other recent determinations also including lattice gauge theory
\cite{gms:97,hoa:98,my:98}. It is $1\,\sigma$ higher than our previous
determination for $m_b$ \cite{jp:97}. The reason was a small inconsistency
in implementing the scale dependence of the $\MSb$ mass in our numerical
program. However, this shift lies well within the range of our estimated
error.

Very recently it has become possible to obtain an estimate of the bottom quark
mass measuring three-jet rates at LEP \cite{rsb:97,delphi:98}. The natural
scale at which $m_b$ is evaluated in this case is $M_Z$. Evolving the value
for $m_b$ determined in our analysis to the $Z$-boson mass, we find:
\begin{displaymath}
\label{eq:mbMz}
m_b(M_Z) \; = \; 2.86 \pm 0.05\,\gev \,.
\end{displaymath}
The result obtained from LEP measurements is at present much less precise
but nevertheless in very good agreement with the value given above. On the
other hand it is the first indication of a scale running of the bottom
quark mass in the $\MSb$ scheme.

Another frequently used mass definition for the bottom quark is the pole
mass. Recent determinations of the bottom quark pole mass include
\cite{vol:95,jp:97,kpp:98,hoa:98}. The pole mass determination from
the $\Upsilon$ system suffers from two difficulties: firstly the
${\cal O}(\as)$ corrections in this case are much larger than for the
$\MSb$ scheme and a resummation of these large Coulombic
corrections has to be performed. The second problem is the fact that the
pole mass suffers from a renormalon ambiguity which introduces an
intrinsic unknown uncertainty.\footnote{For a recent review on renormalons
see \cite{ben:98}.}

Using our result for $m_b$ given above and the perturbative relation between
pole and $\MSb$ mass given in section~2, we obtain $M_b=4.84\pm 0.08\,\gev$.
This result is in agreement with the cited pole mass determinations except
for the previous work by the authors where $M_b=4.60\,\gev$ was found. The
discrepancy is due to the fact that in ref.~\cite{jp:97} the contribution
from the Coulomb poles has been neglected. This contribution formally is of
${\cal O}(\as^3)$ and the analysis \cite{jp:97} was especially aiming for a
consistent treatment of higher orders up to ${\cal O}(\as^2)$.
Nevertheless it turned out that the contribution from Coulomb poles is very
important and should be included. The analysis in the $\MSb$ scheme was not
inflicted by such a problem. We shall come back to these and related questions
in a forthcoming publication.

\bigskip \noindent
{\bf Acknowledgments}
The authors would like to thank S. Narison for the invitation to
this very pleasant and interesting conference.

\end{document}